\documentclass[twocolumn,superscriptaddress,footinbib,pra]{revtex4-1}
\usepackage[utf8]{inputenc}
\usepackage{graphicx}
\usepackage{bm}
\usepackage{amsmath}
\usepackage{amsfonts}
\usepackage{color}

\newcommand{\beq}{\begin{equation}}
\newcommand{\eeq}{\end{equation}}
\newcommand{\beqa}{\begin{eqnarray}}
\newcommand{\eeqa}{\end{eqnarray}}
\def\ra{\rangle}
\def\la{\langle}\def\beq{\begin{equation}}

\begin{document}
\title{Interferometer for force measurement by shortcut-to-adiabatic arms guiding}
\author{A. Rodriguez-Prieto}
\affiliation{Departament of Applied Mathematics, University of the Basque Country UPV/EHU, Bilbao, Spain}

\author{S. Mart\'\i nez-Garaot}
\affiliation{Departament of Physical Chemistry, University of the Basque Country UPV/EHU, Aptdo. 644, Bilbao, Spain}

\author{I. Lizuain}
\affiliation{Departament of Applied Mathematics, University of the Basque Country UPV/EHU, Donostia, Spain}

\author{J. G. Muga}
\affiliation{Departament of Physical Chemistry, University of the Basque Country UPV/EHU, Aptdo. 644, Bilbao, Spain}

\begin{abstract}
We propose a compact  atom interferometer  to measure  homogeneous constant forces guiding the arms via 
shortcuts to adiabatic paths.   
For a given sensitivity, which only depends on the space-time area of the guiding paths, 
the cycle time can be made fast  without loosing visibility.  
The atom is driven by spin-dependent trapping potentials moving in opposite directions, complemented by linear  and time-dependent 
potentials that compensate the trap  acceleration. Thus the arm states are adiabatic in the moving frames, and non-adiabatic in the laboratory frame. 
The trapping potentials may be anharmonic, e.g.  optical lattices, 
and the interferometric phase does not depend on the initial motional state or on the pivot point for swaying the linear potentials.  
\end{abstract}

\maketitle

\section{Introduction}
Atom interferometry \cite{Berman1997,Cronin2009} 
provides a route to quantum-enhanced precise sensors. 
The key idea is to split and later recombine the atom wave function,  to detect the differential phase accumulated during the separation, which is sensitive in particular to small potential differences between the arms. 

Here we work out  a scheme to measure constant forces 
using STA-mediated guided interferometry \cite{Dupont-Nivet2016,Navez2016,Palmero2017,MartinezGaraot2018}. STA stands for ``shortcuts to adiabaticity'', 
a set of techniques to achieve the results of adiabatic dynamics in shorter times \cite{Torrontegui2013,review2019}, and ``guided'' means that  the 
atom is driven in moving traps, as in buckets or conveyor belts \cite{Navez2016},  so it is never in free flight.  
Guiding, e.g. via moving  optical lattices,  keeps the atom wavefunction localized with nanoscale spatial resolution, allowing for 
precise measurements at ultrashort spatial scale \cite{Kovachi2010,Steffen2012}, 
whereas  the speed-up with respect to slow 
adiabatic processes can avoid perturbations and decoherence keeping the visibility and differential phase of adiabatic methods.
Moreover our STA-mediated interferometer scheme fulfills the ideal goal of giving a motional-state independent differential phase 
with short process times while keeping simultaneously a high sensitivity.    

We assume throughout one-particle wavefunctions, either because the interferometer works indeed with single particles or because interactions are negligible. 
Similarly to \cite{MartinezGaraot2018} the arms in the current scheme are separated by means of ``spin'',  here an alias for ``internal state'', dependent forces. 
Operationally the current scheme differs from the one in \cite{MartinezGaraot2018}. There, a fixed harmonic trap was combined  with two  homogeneous  time- and  spin-dependent forces to separate first and then 
recombine the wavefunction branches of an ion.  
Here we use instead two moving spin-dependent traps, not necessarily harmonic, complemented by homogeneous spin- and time-dependent forces to compensate for inertial terms due to the motions of the traps \cite{Torrontegui2011}. This compensation is one of the  
ways to implement STA-driven fast transport \cite{review2019}, and can  be equivalently  found  
by invariant-based inverse engineering
 \cite{Torrontegui2011}, by the   ``fast-forward approach'' \cite{Masuda2010}, or as a local unitary transformation  of a non-local counterdiabatic approach \cite{Ibanez2012,Deffner2014}. 

\begin{figure*}[t!]
\includegraphics[width=125mm]{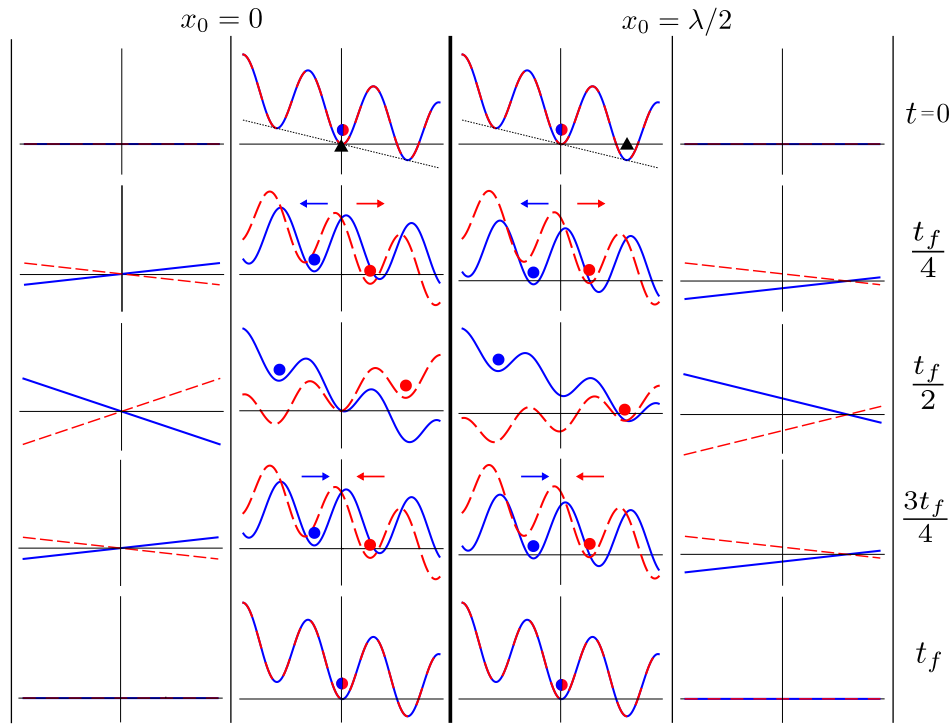}
\caption{Four schematic snapshots of the potentials for moving optical lattices and two choices of pivot $x_0$. 
We use  arbitrary units,
the specific relative values  are not intended to be realistic but are rather chosen to better visualize the process. 
The corresponding $\alpha(t)$ and $f(t)$ are shown in Fig. \ref{figu2}.  $U(x\mp\alpha)=U_0\sin^2(2\pi x/\lambda \mp\alpha)$. The two left columns are for  $x_0=0$ and
the two right columns for $x_{0}=\lambda/2$, a lattice period to the right.
Outer columns: compensating potentials for spin up, $-(x-x_0)f(t)$ (red dashed line)  and  
for spin down $(x-x_0)f(t)$ (blue solid line). In the central columns the triangles are a reminder of  the pivot position and the points  indicate  the moving $\pm\alpha(t)$. The arrows give the sense of motion of the 
lattice.        
$-cx$ (dotted black line) is represented only at $t=0$ but this spin-independent potential acts  throughout the process;  
the wavy lines are the total potentials in Eq. (\ref{HamiltonianUPDOWN}),  ${U}(x-\alpha)-(x-x_0)f(t)-cx$
for spin up (red dashed line) and ${U}(x+\alpha)+(x-x_0)f(t)-cx$ for spin down (solid blue).  
\label{figu1}}
\end{figure*}

An  important difference between this work  and
\cite{MartinezGaraot2018} is that the phase differential is now independent of the pivot,  equipotential point  $x_0$ to apply the compensating spin-dependent potentials, see an example of two different pivot positions in the outer columns of Fig. \ref{figu1}.    
When the force to be measured acts permanently, 
before and during the experiment, the natural choice in which $x_0$ is at the initial equilibrium point of the trap, which depends on the unknown force we want to measure, cancelled the interferometric phase 
in \cite{MartinezGaraot2018}.
A  rotation of the effectively one-dimensional (1D) trap to let the force act only from the initial time $t=0$ is a formal, but hardly practical  solution. The scheme proposed here is free from such difficulties and is also  more broadly applicable.

Using  arbitrary trapping potentials, rather than harmonic ones,  opens the way to applying the proposed scheme  
to ultracold neutral atoms where the anharmonicities are usually stronger than for trapped ions. Different  realizations are possible, 
e.g. in atom chips \cite{Cronin2009}, but we shall discuss optical lattices as a specific example.  Interferometers with 
two oppositely moving optical lattices to accelerate the  arm wavefunctions  for a single internal state have been 
demonstrated \cite{Muller2009} and studied theoretically \cite{Kovachi2010}. Closer to our goal,  
Mandel et al. \cite{Mandel2003} demonstrated   transport  of the spin-dependent wavefunctions in optical lattices moving in opposite directions 
with a scheme proposed in \cite{Jaksch1999}, 
and  Steffen et al. \cite{Steffen2012} built a single-atom  interferometer based on a similar setting.   To implement our scheme we envision,  for each spin, double supperlattices composed by an ultra-deep optical lattice as a ``conveyor-belt'' trapping potential \cite{Schrader2001} with negligible tunneling,  see \cite{Lu2020} and references therein, while the  compensating force may be achieved by a second lattice with much larger periodicity
than the trapping lattice to make it effectively homogeneous for each arm wavefunction. 
Factors of ten between the periodicities of two lattices are routinely found playing with the angle 
between the crossing beams \cite{Williams2008} and even higher factors are technically possible \cite{Tao2018}. 

First we present the main idea of the interferometer and basic relations in Sec. \ref{sec:interferometer} and then the 
recipe to move the arm  traps and set the time dependence of the compensating forces in Sec. \ref{sec:Phases}.
The theory relies on a transformation to  ``moving-frame'' interaction pictures for each arm. An alternative formulation is presented in 
Sec. \ref{sec:invariants}  
in terms of ``invariants'' which connects the current approach to ``invariant-based inverse engineering'' \cite{Torrontegui2011,Palmero2017,MartinezGaraot2018}. The interferometric phase can then be simply interpreted as the difference between the Lewis-Riesenfeld phases for the arms. This connection enables us to use invariant-based concepts and results, for example to apply techniques to enhance robustness with respect to different noises 
\cite{Ruschhaupt2012,Lu2018,Lu2020}.
The paper ends with a discussion on possible applications and open questions. 
\section{Basic concept of  the interferometer}
\label{sec:interferometer}
Consider an atom  with two internal states, ``spin up'' $|\!\uparrow\rangle$, and ``spin down'' $|\downarrow\rangle$,
and effective motion in one dimension.  A general state at time $t$ is 
$
a_{\uparrow} |\!\uparrow \rangle \psi^{\uparrow}(x,t) + a_{\downarrow}  |\!\downarrow \rangle
\psi^{\downarrow}(x,t),
$
where $\psi^{\uparrow}(x,t)=\langle x|\psi^\uparrow(t)\rangle$ and $\psi^{\downarrow}(x,t)=\langle x|\psi^\downarrow(t)\rangle$ are the motional states for the two internal levels, in coordinate representation. We assume a prepared state  $|\!\uparrow\rangle |\Phi_p\ra$
from which  
a $\pi/2$ pulse   \cite{Leibfried2003} produces two equally weighted components with $a_{\uparrow}=a_{\downarrow}=1/2^{1/2}$.
We set time $t=0$ at the end of the $\pi/2$ pulse and, assuming a Lamb-Dicke regime and a fast pulse compared to motional periods,   
$\Phi(x,0)\equiv\Phi_{p}(x)=\psi^{\uparrow}(x,0)=\psi^{\downarrow}(x,0)$. 
The two branches evolve separately in coordinate space due to spin-dependent forces.
At some final time $t_f$ the complex overlap can be written in polar form as 
\beq
\label{o}
\langle \psi^{\downarrow}(t_f)|\psi^{\uparrow}(t_f) \rangle=e^{i\Delta\varphi(t_f)}|\langle \psi^{\downarrow}(t_f)|\psi^{\uparrow}(t_f) \rangle|, 
\eeq
A second $\pi/2$ pulse 
gives the populations \cite{MartinezGaraot2018} 
\beq
P^{\uparrow\downarrow}(t_f)= \frac{1}{2}\pm \frac{1}{2}\Re{\rm e} \left [ \langle \psi^{\downarrow}(t_f)|\psi^{\uparrow}(t_f) \rangle \right ],
\eeq
where we have neglected the $\pi/2$-pulse duration. 
The STA-driving will achieve  maximal visibility, i.e., $|\langle \psi^{\downarrow}(t_f)|\psi^{\uparrow}(t_f)\rangle|=1$, note that  $|\langle \psi^{\uparrow\downarrow}(t_f)|\Phi(0)\rangle|=1$ is {\it not} required.  
Then the populations read
$
P^{\uparrow\downarrow}(t_{f})=\frac{1}{2}\pm\frac{1}{2}\cos[\Delta\varphi(t_{f})]. 
$
If the differential phase is proportional to a constant force $c$, $\Delta\varphi(t_{f})={\cal S}c$
and the sensitivity ${\cal S}$ is known,
$c$ can be measured 
from the populations.
When   
$c$, or its deviation from some approximately known value, are expected to be small in the scale of  $\pi/{\cal S}$, $c$ is found from the populations using the relevant branch of the arccosine.  
More generally, $c$ may be found unambiguously from the periodicity ${2\pi}/{c}$
of the populations
$P^{\uparrow\downarrow}(t_{f})$ 
as a function of ${\cal S}$   \cite{MartinezGaraot2018}. Measuring the populations requires repetitions in time if the interferometer works with a single particle, or alternatively noninteracting ensembles.   

The method to guide the arm wave functions  described below  
fulfills the  hypotheses made so far, namely, the modulus of the overlap (\ref{o}) is one and the differential phase is proportional to $c$. Moreover it will be possible to control the sensitivity ${\cal S}$ and the time $t_f$ of the process independently of $|\Phi(0)\rangle$.  

%
%%%%%%%%%%%%%%%%%%%%%%%%%%%%%%%%%%%%%%%%%%%%%%%%%%%%%%%%%%%%%
%%%%%%%%%%%%%%%%%%%%%%%%%%%%%%%%%%%%%%%%%%%%%%%%%%%%%%%%%%%
%%%%%%%%%%%%%%%%%%%%%%%%%%%%%%%%%%%%%%%%%%%%%%%%%%%%%%%%%%%%%
\begin{figure}[t!]
\includegraphics[width=78mm]{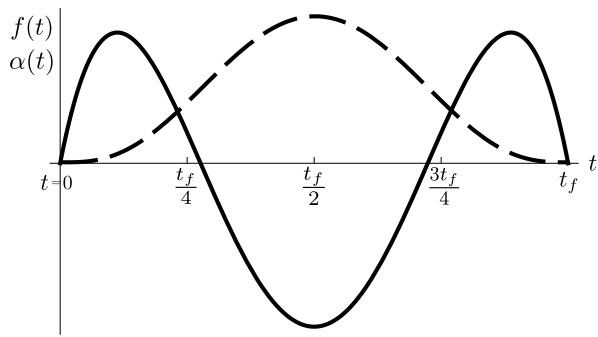}
\caption{Typical  forms of  $\alpha(t)$ (dashed line),  and  $f(t)$  (solid line) in arbitrary units:  $\alpha(t)$ is found by imposing 
the boundary conditions (\ref{bc_y}) and $\ddot{\alpha}(t_b)=0$ at times $t_b=0,t_f$  to a polynomial  ansatz
\cite{MartinezGaraot2018}, see Eq. (\ref{bs}). 
$f(t)$ is found from $\alpha(t)$  via Eq. (\ref{Newt}), it vanishes at the boundary times and integrates to zero. 
  \label{figu2}}
\end{figure}
%%%%%%%%%%%%%%%%%%%%%%%%%%%%%%%%%%%%%%%%%%%%%%%%%%%%%%%%%%%
%
%
%
%
%
\section{How to move the guiding traps}
\label{sec:Phases}
For each spin state we assume a different evolution driven by 
the  Hamiltonians  (Here and in the following, the superscript $\uparrow\downarrow$ in any equation implies that  the sign on top in $\mp$ or $\pm$ is for  $\uparrow$, whereas the sign on the bottom is for $\downarrow$.)
\begin{equation}
\label{HamiltonianUPDOWN}
H^{\uparrow\downarrow}=\frac{p^{2}}{2m}- cx\mp \left[x-x_{0}(t)\right] f(t)+U[x\mp \alpha(t)].
\end{equation}
The trap potentials ${U}[x \mp \alpha(t)]$ 
move along opposite trajectories  $\alpha^{\uparrow\downarrow}(t)=\pm\alpha(t)$.
We consider trap trajectories  that satisfy the boundary conditions 
\beq
\alpha(t_{b})=\dot \alpha(t_{b})=0
\label{bc_y}
\eeq
at the boundary times $t_{b}=0, t_{f}$. The dots stand for time derivatives. 
Each trap starts and ends at rest returning to the starting point, equal for both traps. 

The trap potentials are complemented by spin-dependent linear potentials 
$\mp \left[x-x_{0}(t)\right]f(t)$ that cross at the pivot point $x_0(t)$:  In a typical experiment $x_0$ will be 
constant, however we shall  
keep by now formally a more general  $x_0(t)$.
The force $f(t)$ will be chosen to compensate inertial terms in the moving frame as discussed below in detail. 
Finally,  
$c$ is the spin-independent  homogeneous-in-space and constant-in-time force  that we want to measure,  
$m$ is the mass of the atom, and ${p^{2}}/{2m}$ the kinetic energy.  Examples of the potential terms in 
Eq. (\ref{HamiltonianUPDOWN})  
are depicted schematically in Fig. \ref{figu1} for  $U$ as an optical lattice potential and for two different  pivots.  

Let us now reorganize the Hamiltonians (\ref{HamiltonianUPDOWN}) as follows, 
\begin{eqnarray}
H^{\uparrow\downarrow}
= \frac{p^{2}}{2m}\mp f(t) x+\widetilde{U}(x\mp\alpha)+\Lambda^{\uparrow\downarrow}(t), 
\eeqa
where we have separated purely time-dependent terms in 
\beq
\Lambda^{\uparrow\downarrow}(t)=\pm f(t)x_{0}(t) \mp c\alpha(t)
\eeq
and defined effective trap potentials 
$\widetilde{U}$ that include the effect of the force $c$, 
\begin{eqnarray}
\widetilde{U}(x\mp \alpha(t))=U(x\mp \alpha(t))- [x \mp  \alpha(t)]c.
\end{eqnarray}
To solve the dynamics, it is useful to perform   unitary transformations into 
``moving-frame interaction pictures''. 
Specifically we define the interaction picture wavevectors $|\psi_I^{\uparrow\downarrow}\rangle$ in terms of 
Schr\"odinger (laboratory frame) wavevectors  $|\psi^{\uparrow\downarrow}\rangle$,   
as
\beq
|\psi_I^{\uparrow\downarrow}\rangle={\cal U}^{\uparrow\downarrow}|\psi^{\uparrow\downarrow}\rangle,\;\;\;\;\; 
|\psi^{\uparrow\downarrow}\rangle=({\cal U}^{\uparrow\downarrow})^\dagger |\psi_I^{\uparrow\downarrow}\rangle,
\label{IP}
\eeq
where the unitary operator ${\cal U}^{\uparrow\downarrow}$ is constructed by multiplying position and momentum 
shift operators, 
\beq
{\cal U}^{\uparrow\downarrow}=e^{\pm i\alpha p/\hbar}e^{\mp i m \dot{\alpha} x/\hbar}.
\label{order}
\eeq
Other orderings and therefore interaction pictures are possible but the measurable quantities and the 
differential phase are not affected by the different orderings as long as the intermediate calculations are done consistently. Using Eq. (\ref{order})    
for each arm, the effective, moving-frame Hamiltonians become 
\beqa
H_I^{\uparrow\downarrow}&=&{\cal U}^{\uparrow\downarrow}H^{\uparrow\downarrow}({\cal U}^{\uparrow\downarrow})^\dagger+i\hbar\,\dot{{\cal U}}^{\uparrow\downarrow}({\cal U}^{\uparrow\downarrow})^\dagger
\nonumber\\
&=&\frac{p^2}{2m}+\frac{1}{2}m\dot{\alpha}^2\mp (x\pm\alpha)f(t)+\widetilde{U}(x)
\nonumber\\
&\pm& f(t) x_0(t)\mp c\alpha \pm (x\pm \alpha)m\ddot{\alpha}.
\label{hami0}
\end{eqnarray}
If the applied $f(t)$ satisfies  
\begin{equation}
\ddot  \alpha(t)=\frac{f(t)}{m},
\label{Newt}
\end{equation}
which can be interpreted as a Newton equation for the trajectory $\alpha(t)$ subjected to the force $f(t)$, 
this auxiliary force compensates for inertial effects due to the motion of the 
$\widetilde{U}(x\mp \alpha(t))$ potentials in the laboratory frame, note the cancellation of the third and last terms in Eq. (\ref{hami0}). The consequence is that 
a stationary state in the moving frame will remain so. Equation (\ref{Newt}) is used inversely, i.e., $f(t)$ is found from a designed $\alpha(t)$ and hereinafter $f(t)$ is always assumed to satisfy Eq. (\ref{Newt}), {except in point { vii} of the final discussion}.  
To make $f(t_b)$ zero at the boundary times we shall impose, in addition to the boundary conditions (\ref{bc_y}), that 
$
\ddot{\alpha}(t_b)=0.
$
Applying Eq. (\ref{Newt}) in Eq. (\ref{hami0}) the moving-frame Hamiltonians take a simple form with a common time- and spin-independent term $H_{I,0}$, and terms $F^{\uparrow\downarrow}(t)$ that depend on time, but not on $x$ or $p$, 
\begin{eqnarray}
H_I^{\uparrow\downarrow}&=&H_{I,0}+F^{\uparrow\downarrow}(t),\;
H_{I,0}=\frac{p^2}{2m}+\widetilde{U}(x),
\nonumber\\
F^{\uparrow\downarrow}(t)&=&\frac{1}{2}m\dot{\alpha}^2\pm f(t) x_0(t) \mp c \alpha(t).  
\label{structure}
\end{eqnarray}
The resulting structure facilitates the formal solution of the dynamics,
as the time-dependent part only accumulates a phase, 
whereas the time-independent part gives a simple evolution operator, 
\beqa
|\psi_I^{\uparrow\downarrow}(t)\rangle &=&e^{- i\int_0^t F^{\uparrow\downarrow}(t')dt'/\hbar}
|\psi_{I,0}^{\uparrow\downarrow}(t)\rangle,
\nonumber\\
|\psi_{I,0}^{\uparrow\downarrow}(t)\rangle&=&e^{-iH_{I,0}t/\hbar}|\psi_{I,0}^{\uparrow\downarrow}(0)\rangle.
\eeqa
As $|\psi_{I,0}^{\uparrow\downarrow}(0)\rangle=|\Phi(0)\rangle$ and $H_{I,0}$ are spin independent,   
$|\psi_{I,0}^{\uparrow\downarrow}(t)\rangle=|\Phi(t)\rangle=e^{-iH_{I,0}t/\hbar}|\Phi(0)\rangle$ is also  a spin-independent vector.     

Noting that  $e^{\mp i \alpha p/\hbar}$ shifts the position representation as   
$
\langle x| e^{\mp i \alpha p/\hbar}|\Phi\rangle= \Phi(x\mp\alpha),
$
the branch wave functions in the laboratory frame are found by Eq. (\ref{IP}), 
\begin{equation}
\psi^{\uparrow\downarrow}(x,t)=e^{\pm im\dot{\alpha} x/\hbar} e^{- i\int_0^t\! F^{\uparrow\downarrow}(t')dt'/\hbar}
\Phi(x\mp\alpha,t). 
\label{wfs}
\end{equation}
In particular, at final time $t_f$, 
\begin{eqnarray}
\psi^{\uparrow\downarrow}(x,t_f)&=&e^{-im\!\int_0^{t_f}\! \dot{\alpha}^2 dt/2\hbar} e^{\pm i c\!\int_0^{t_f}\! \alpha(t) dt/\hbar} 
\nonumber\\
&\times&e^{\mp i\! \int_0^{t_f}\! x_0(t)f(t) dt/\hbar} \Phi(x,t_f). 
\label{wfsud}
\end{eqnarray}
For $x_0$ constant but otherwise arbitrary,  the overlap in Eq. (\ref{o}) takes a very simple form, since  the phase terms $\mp x_0\int_0^{t_f} f(t) dt=0$ vanish because of Eq. (\ref{Newt})  and the 
boundary condition $\dot{\alpha}(t_b)=0$,
\begin{equation}
\langle \psi^{\downarrow}(t_f)|\psi^{\uparrow}(t_f) \rangle=e^{2 i c\int_0^{t_f}\! \alpha(t) dt/\hbar},
\label{simple}
\end{equation}
so that $c$ can be measured from the interferometric differential phase via the 
populations as explained before.
The phase is indeed 
proportional to $c$, 
$
\Delta\varphi(t_f)= c{\cal S},
$
with controllable  
sensitivity
\beq
{\cal S}=\frac{2}{\hbar} \int_0^{t_f}\! \alpha(t) dt, 
\label{sensi}
\eeq
the  space-time area $2\int_0^{t_f} \alpha(t) dt$ in units of $\hbar$ swept between the two trap paths. Because the relative motion of the motional states with respect to $\pm \alpha(t)$ is identical in both arms, 
this is the same area between  the state centroids for any initial motional state $|\Phi(0)\ra$.
Thus the interferometric phase and sensitivity are independent of the initial motional state, a robust 
 ``geometrical'' feature of the 
proposed  interferometer.     
  
$t_f$ can be chosen freely, in particular it can be made short 
compared to relevant decoherence times, and $\alpha(t)$ can be manipulated 
to change the sensitivity. Examples on how to set $\alpha(t)$  may be found in \cite{MartinezGaraot2018}, the basic idea 
is to expand it in some basis,  e.g. sines or powers of $t$,  with enough number of terms to satisfy the boundary conditions.
More terms are added if further conditions are imposed, such as a desired value of ${\cal S}$.   
\subsection{Example: Sensitivity for Caesium atom interferometer} 
\begin{figure}[t!]
\includegraphics[width=85mm]{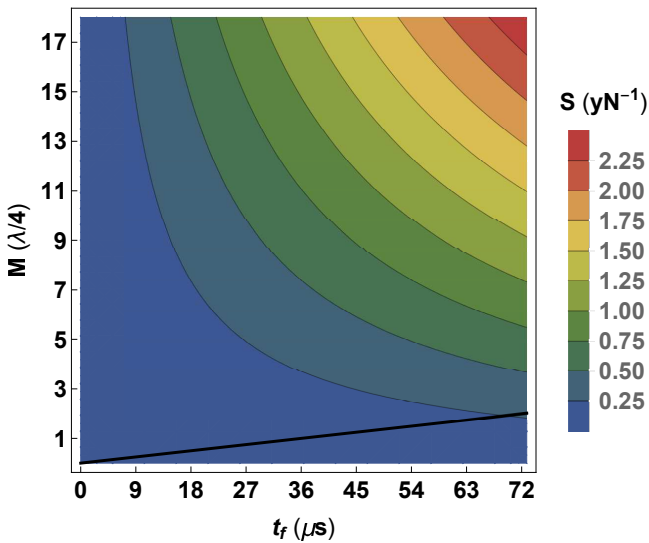}
\caption{Contour plot of the sensitivity ${\cal S}=\frac{32 M t_f}{35 \hbar}$ found for $\alpha(t)$ in Eq. (\ref{bs}), see also Fig. \ref{figu2}.
The solid black line is a linear dependence of $M=\frac{\lambda}{72 \mu s} \frac{t_f}{2}$  that extrapolates the one applied in  \cite{Steffen2012} for an elementary displacement.  
$\lambda=866$ nm. \label{figu3}}
\end{figure}

Steffen et al. \cite{Steffen2012} implemented a Caesium atom interferometer which demonstrates some 
elements of the current scheme, specifically the atom wavefunction was split into separated paths 
controlled by spin-dependent optical-lattice potentials. A large displacement of $\pm\alpha(t)$ was technically -not fundamentally-  limited by the maximum voltage that can be applied to an 
electro-optical modulator 
\cite{Mandel2003,Steffen2012}. This limitation was circumvented  by accumulating elementary operation blocks 
which consist of lattice displacements with alternating direction interleaved by $\pi$-pulses  
\cite{Mandel2003,Steffen2012}. A single lattice displacement  by $\lambda/4$ took 18 $\mu$s. %An idealized  
%trap trajectory with the corresponding velocity $\lambda/(72 \mu s)$ 
%is plotted as a straight line in Fig. ... 
Figure \ref{figu3} represents a contour plot of the sensitivities (\ref{sensi}) 
with a  four-term polynomial $\alpha(t)$ satisfying the boundary conditions  (\ref{bc_y}) and $\ddot{\alpha}(t_b)=0$, with its maximum value $M$ at $t_f/2$,  see Fig. \ref{figu2}, 
%We have used sixth order polynomials to design a first type of trajectory
%
\beqa
%\label{alphaA}
\alpha(t)&=&\sum_{j=3}^{6} b_{j} \left(\frac{t}{t_{f}}\right)^{j}, 
\nonumber\\
%\end{equation}
%
%where 
%
%\beqa
b_3&=&64 M,\;\;
b_4=-192 M,
\nonumber\\
b_5&=&192 M,\;\;
b_6=-64 M.\label{bs}
\eeqa
%
%This is the $\alpha(t)$ shown in Fig. \ref{figu2}. 
The resulting sensitivity ${\cal S}$ is remarkably simple, namely, ${\cal S}=32 M t_f/(35\hbar)$. 
Note that the scaling of ${\cal S}$ 
with $t_f$ can be chosen at will by fixing the  dependence of $M$ on $t_f$, this amounts to follow a line $M(t_f)$ in Fig. \ref{figu3}, for example
as $M\propto t_f^\mu$, with $\mu=0,1,2,...$. Assuming a dependence of the order of the elementary displacement in \cite{Steffen2012},
gives $M=\frac{\lambda}{72 \mu s}\frac{t_f}{2}$, see the straight line in Fig. \ref{figu3}, 
and an ${\cal S}$ that depends quadratically on $t_f$. With the current scheme the trap can be subjected to  strong
accelerations without spoiling the visibility since they are compensated. Thus, for a given $t_f$  higher sensitivities can be achieved for faster dependences $M(t_f)$. 
Formally there is no limit to how large ${\cal S}(t_f)$ may be. The limit will be set in practice by the 
technical limitations imposed by the specific setting to implement  $\alpha(t)$ and $f(t)$. For a given, desired sensitivity ${\cal S}_0$, $M(t_f;{\cal S}_0)$ depends, along a given contour in Fig. \ref{figu3},   inversely on $t_f$, 
$M(t_f;{\cal S}_0)=35 {\cal S}_0\hbar/(32 t_f)$. If $M$ is technically limited by some upper value, $t_f$ will be lower limited accordingly.            

\section{Invariants}
\label{sec:invariants}
In this section we will connect the results found so far  with 
Lewis-Riesenfeld invariants of motion 
and the inverse engineering of trap trajectories based on them 
\cite{Torrontegui2011,Torrontegui2013}.  
%[M7,M11].
%Equation numbers (..) and reference numbers [...] from the main text are cited here preceded by ``M''. 

A  key result is the moving-frame Hamiltonian structure found in 
Eq. (\ref{structure}). The Hamiltonian $H_{I,0}$ does not depend on time, and therefore  
its expectation value $\langle\Phi(t)|H_{I,0}|\Phi(t)\rangle$ is constant. In the laboratory frame, making use of Eq. (\ref{IP}) 
this translates into 
\beq
\frac{d}{dt}\langle \psi^{\uparrow\downarrow}(t)|({\cal U}^{\uparrow\downarrow})^\dagger H_{0,I}{\cal U}^{\uparrow\downarrow}|\psi^{\uparrow\downarrow}(t)\rangle=0,
\label{invh0}
\eeq
or, in other words, 
\beq
I^{\uparrow\downarrow}\equiv({\cal U}^{\uparrow\downarrow})^\dagger H_{0,I}{\cal U}^{\uparrow\downarrow}
\eeq
are ``dynamical'' Lewis-Riesenfeld invariants of motion for, respectively, the branch Hamiltonians  
$H^{\uparrow\downarrow}$  in Eq. (\ref{HamiltonianUPDOWN}),  supplemented  by Eq. (\ref{Newt}) to specify $f(t)$. They satisfy the invariance equations 
\begin{equation}
 \frac{dI^{\uparrow\downarrow}}{dt}=\frac{\partial I^{\uparrow\downarrow}}{\partial t} + \frac{1}{i\hbar}\left[I^{\uparrow\downarrow}, H^{\uparrow\downarrow}\right]. 
\end{equation}
These invariants may be  calculated explicitly with the aid of  Eq. (\ref{order}),
\beq
I^{\uparrow\downarrow}=\frac{1}{2m}(p\mp m\dot{\alpha})^2+\widetilde{U}(x\mp\alpha),
\eeq
and their (constant-in-time) eigenvalues $\lambda_n$ are nothing but the eigenvalues of $H_{I,0}$,
\begin{equation}
\label{statio}
 \left[\frac{-\hbar^{2}}{2m}\frac{\partial^{2}}{\partial x^{2}}+\widetilde{U}(x)\right]\phi_{n}(x)=\lambda_{n}\phi_{n}(x),
\end{equation}
where the  $\phi_n(x)$ are the eigenfuctions of $H_{I,0}$. They form a natural basis to expand $|\Phi(t)\rangle$ as 
\beqa
|\Phi(t)\rangle&=&\sum_n e^{-i\lambda_n t/\hbar} |\phi_n\rangle c_n,
\nonumber\\
c_n&=&\langle \phi_n|\Phi(0)\rangle,
\label{superpo}
\eeqa
in terms of constant coefficients $c_n$. 
The vectors 
\beq
|\psi_n^{\uparrow\downarrow}\rangle\equiv ({\cal U}^{\uparrow\downarrow})^\dagger|\phi_n\rangle
\label{psin}
\eeq
are eigenvectors of $I^{\uparrow\downarrow}$ with eigenvalue $\lambda_n$ since 
\beqa
I^{\uparrow\downarrow}|\psi_n^{\uparrow\downarrow}\rangle&=&({\cal U}^{\uparrow\downarrow})^\dagger H_{I,0}\,{\cal U}^{\uparrow\downarrow}({\cal U}^{\uparrow\downarrow})^\dagger|\phi_n\rangle
\nonumber\\
&=&\lambda_n ({\cal U}^{\uparrow\downarrow})^\dagger |\phi_n\rangle=\lambda_n|\psi_n^{\uparrow\downarrow}\rangle.
\eeqa
Using the explicit form of ${\cal U}^{\uparrow\downarrow}$ in Eq. (\ref{order})
their coordinate representation is 
\beq
\psi_n^{\uparrow\downarrow}(x,t)=e^{\pm i m\dot{\alpha}x/\hbar}\phi_n(x\mp \alpha).
\eeq
The ``dynamical modes'' are defined as orthogonal solutions of the time-dependent Schr\"odinger equations driven by $H^{\uparrow\downarrow}$ proportional to 
these eigenfunctions, $e^{i\theta_n^{\uparrow\downarrow}(t)}\psi_n(x,t)$, where the Lewis-Riesenfeld phases $\theta_n^{\uparrow\downarrow}(t)$
are found from    
\begin{equation}
\frac{d \theta^{\uparrow\downarrow}_{n}(t)}{d t}=\frac{1}{\hbar}
\left\langle\psi^{\uparrow\downarrow}_{n}(t)\left|i\hbar\frac{\partial}{\partial t}-H^{\uparrow\downarrow}\right|\psi^{\uparrow\downarrow}_{n}(t)\right\rangle,
\label{previous}
\end{equation}
so that the Schr\"odinger equations are satisfied. 
Setting $\theta_n^{\uparrow\downarrow}(0)=0$, an explicit calculation gives, see the Appendix, 
\beq
\theta^{\uparrow\downarrow}_{n}(t)=\frac{-1}{\hbar}\int_{0}^{t}\left[\lambda_{n}+F^{\uparrow\downarrow}(t')\right]dt'.
\label{LR}
\eeq
Arbitrary wavefunction solutions of the dynamics $\psi^{\uparrow\downarrow}(t)$   will combine 
these elementary solutions with constant coefficients. For the initial state $|\Phi(0)\rangle$, 
$\psi^{\uparrow\downarrow}(x,t)=\sum_n e^{i\theta_n^{\uparrow\downarrow}(t)}\psi_n(x,t) c_n$. 
Factoring out $n$-independent 
phase factors and summing over $n$ as in Eq. (\ref{superpo}), 
the expression (\ref{wfs}) and following results for  $\psi^{\uparrow\downarrow}(x,t)$ in the main text are exactly recovered. 
The interferometric phase from this point of view is nothing but the  difference between Lewis-Riesenfeld phases for the arms. 
Since the $n$-dependent part cancels out, the result is $n$-independent.     

\section{Discussion \label{final}}
We have put forward an STA-mediated atomic  interferometer scheme to measure homogeneous constant forces with spin- (internal-state) dependent moving traps to guide the wavefunction components along the two arms.    
The approach is robust in different ways:  

i) As the process can be made fast,  
decoherence effects and perturbations can be mitigated or avoided without necessarily
renouncing to some required sensitivity. For a caveat on the relation between process time and decoherence see point vi below.   

ii) The moving trapping potentials may be anharmonic, so the method  may be applied in particular to optical lattices 
as conveyor belts to drive the arms.   

iii) The motional initial wave function is arbitrary, there is no need to prepare a perfect ground state because the differential 
phase is not affected by the initial motional state.

iv) The moving trap potentials are complemented by time-dependent linear potentials that compensate inertial forces  
``rocking''  on a pivot point $x_0$.  
The differential interferometric phase is simplified and made pivot independent  when  
$
\int_0^{t_f}\! x_0(t)f(t) dt=0
$
in Eq. (\ref{wfsud}). The integral vanishes  when  $x_0$ is a constant because of the way $f(t)$ is constructed. 
This result is in fact robust with respect to  typical forms of $x_0(t)$: A noisy $x_0(t)$ with a zero-mean perturbation around its nominal constant value
will give a vanishing integral  as long as the correlation time is short  compared to 
$t_f$. Other relevant dependence is an  undesired linear drift, e.g. $x_0(t)=a +bt$. For the linear term $bt$ 
integrating by parts and using  the boundary conditions (\ref{bc_y}) gives a zero integral too. 
It may also be of interest in practice to set a spin-dependent $x_0(t)$. For example, for  $x_0^{\uparrow\downarrow}(t)=\pm\alpha(t)$ the resulting integrals $\pm\int_0^{t_f} [\pm\alpha(t)]f(t) dt$ would not vanish but they would  give the same phase for both arms, which makes  the differential phase again pivot independent.    
      
v) There is ample freedom to choose the  trap paths $\pm\alpha(t)$ which are only subjected, apart from technical limitations,  to 
satisfy some boundary conditions
at initial and final  times. This flexibility may be used to  change the sensitivity ${\cal S}$.  It also allows to achieve fast scalings of ${\cal S}$ 
with total time $t_f$, in principle with an arbitrary power of $t_f$, 
in contrast to linear scaling with $t_f$ of Ramsey-Bord\'e interferometers or with $t_f^2$ in a Mach-Zender configuration 
\cite{McDonald2014}.

vi) Following techniques developed to enhance the robustness of  STA approaches \cite{Ruschhaupt2012,review2019}, 
the freedom in choosing $\alpha(t)$ may be used to make the differential phase  robust against specific setting-dependent perturbations,  e.g. some particular type of noise relevant for the experimental arrangement. A recent study \cite{Lu2020} analyzes the motional energy excitation of atoms due to noises affecting different moving optical lattice  parameters: periodicity, depth, or position. The excitations may be 
analyzed in terms of static or dynamical contributions whose relative importance depends on the parameter affected by noise. 

Static contributions are defined as those which are independent of the trap trajectory,  they just increase with transport time $t_f$ so the strategy to mitigate them is to shorten process times. They are dominant in particular for position noise.  

Dynamical contributions depend on the 
trap trajectory so they could be mitigated by a good choice of $\alpha(t)$. For  ``accordion noise'' of the lattice periodicity they 
dominate and give  minimal excitation at a certain transport time. For noise in the trap depth there is also  a time $t_f$ with minimal excitation 
 with dynamical terms dominating at shorter transport  times and static terms at larger times.

The existence of minima -for some but not all noise types- underlines  that the naive expectation that   
shorter and shorter times $t_f$ are always beneficial is not necessarily correct. The beneficial effect of shortening the time depends on the noise type and on the time domain. 
It also points out that there are no universal recipes, 
each noise or perturbation requires a dedicated study. Adapting the analysis in \cite{Lu2020}, which did not include compensating forces,  to the current configuration, is left for a separate work.  
   
vii) The arm wavefunctions overlap and differential phase found in Eq. (\ref{simple}) are exact, i.e., no adiabatic approximation has been performed, and there is no need to calculate  non-adiabatic corrections. In this regard it is interesting to sketch how this result is found in 
the adiabatic, slow motion limit when the compensating force $f(t)$ is {\it not} applied. The calculation would start in Eq. (\ref{hami0}) for 
$H_{I}^{\uparrow\downarrow}$.  Taking now $f(t)=0$ these moving-frame Hamiltonians  cannot be separated into purely $t$-independent  and purely $t$-independent terms because  of the inertial terms $\pm xm\ddot\alpha$. In the slow-motion limit, however, these terms will be  negligible compared to 
$\widetilde{U}$  so that 
the structure in that limit is again that of a time-independent Hamiltonian and purely time-dependent terms. The corresponding dynamics then lead to Eq. (\ref{simple}), but  only as an approximation. In contrast, when the compensation forces $f(t)=\mp m\ddot{\alpha}$ 
are applied, the dynamics is generally non-adiabatic in the laboratory frame, but adiabatic by construction in the moving frames, a key property that allows us to set 
simultaneously short process times and large sensitivities.   
         
We hope that the unique features of the proposed scheme, among them independence of initial state, arbitrary trap potential, 
and freedom to choose sensitivity and cycle time, will motivate further work. 
The elements necessary to implement the current scheme have been separately demonstrated.
%A detailed analysis of the technical challenges implied is beyond the scope of 
%this work, but we note that in  realizations of oppositely moving spin-dependent lattices, a large displacement of $\pm\alpha(t)$ was technically -not fundamentally-  limited by the maximum voltage that can be applied to an 
%electro-optical modulator 
%\cite{Mandel2003,Steffen2012}. This limitation was circumvented  by accumulating elementary operation blocks 
%which consist of lattice displacements with alternating direction interleaved by $\pi$-pulses  
%\cite{Mandel2003,Steffen2012}.  
We have paid some attention to the use of oppositely moving spin-dependent optical lattices \cite{Mandel2003,Steffen2012}. 
Alternative realizations may be based on the unitary equivalence between the ``local'', position- and $t$-dependent
compensating Hamiltonian terms
$\mp m\ddot{\alpha}(t)x$  and ``counterdiabatic'' momentum- and $t$-dependent terms $\pm p\dot{\alpha}$ \cite{Ibanez2012,review2019}. 
While implementing the former in the laboratory is quite generally easier than the latter, the spin-dependent counterdiabatic terms may be realized in systems with either actual or synthetic spin-orbit coupling \cite{Ban2012,Cadez2013,Cadez2014,Chen2018}.

\acknowledgments{We thank D. Leibfried, J. Bollinger, and D. Gu\'ery-Odelin for many useful discussions. 
This work was supported by the Basque Country Government (Grant No. IT986-16), and 
by   PGC2018-101355-B-I00 (MCIU/AEI/FEDER,UE). }

\appendix
\section{Calculation of Eq. (\ref{LR})\label{app2}}
To calculate the Lewis-Riesenfeld phases in Eq. (\ref{LR}) we start from calculating the matrix elements in 
Eq. (\ref{previous}).  
It proves convenient to  write first, using Eqs. (\ref{hami0}) and ({\ref{structure}),  
\beq 
H^{\uparrow\downarrow}=I^{\uparrow\downarrow}+F^{\uparrow\downarrow}-i\hbar({\cal U}^{\uparrow\downarrow})^\dagger
\dot{\cal U}^{\uparrow\downarrow}.
\eeq
Using now Eq. (\ref{psin})
we find that  
\beqa
&-&\la \psi^{\uparrow\downarrow}_n|H^{\uparrow\downarrow}| \psi^{\uparrow\downarrow}_n\ra/\hbar
\nonumber\\
&=&\frac{-1}{\hbar}(\lambda_n+F^{\uparrow\downarrow})+i\la \phi_n| \dot{\cal U}^{\uparrow\downarrow}({\cal U}^{\uparrow\downarrow})^\dagger
|\phi_n\ra
\label{first}
\eeqa
whereas, using again   Eq. (\ref{psin}) and noting that $\dot{\cal U}^{\uparrow\downarrow}({\cal U}^{\uparrow\downarrow})^\dagger=-{\cal U}^{\uparrow\downarrow}(\dot{\cal U}^{\uparrow\downarrow})^\dagger$,   
\beq
i\la \psi_n^{\uparrow\downarrow}|\dot{\psi}_n^{\uparrow\downarrow}\ra=-i \la \phi_n| \dot{\cal U}^{\uparrow\downarrow}({\cal U}^{\uparrow\downarrow})^\dagger|\phi_n\ra.
\label{second}
\eeq
The right hand side  may be calculated explicitly  but in any case it is cancelled by  the last term in Eq. (\ref{first}) 
when summing  Eqs. (\ref{first}) and (\ref{second}) in Eq. (\ref{previous}). Integrating we get   
finally Eq. (\ref{LR}).  

\end{document}